\documentclass[preprint,preprintnumbers,amsmath,amssymb]{revtex4}
\usepackage{graphicx}
\usepackage{dcolumn}
\usepackage{bm}
\usepackage{epsfig}

\def\integral{\int_{-\infty}^{+\infty}}
\def\pre{\frac{1}{\sqrt{2\pi\hbar}}}
\begin{document}

\preprint{\bf PSU/TH-253} 


\title{Analytic results for Gaussian wave packets in four model systems:
I. Visualization of the kinetic energy}

\author{R. W. Robinett and L. C. Bassett}
\affiliation{
Department of Physics\\
The Pennsylvania State University\\
University Park, PA 16802 USA \\
}

\date{\today}

\begin{abstract}
Using Gaussian wave packet solutions, we examine how the kinetic energy 
is distributed in time-dependent solutions of the Schr\"odinger
equation corresponding to the cases of a free particle, a particle undergoing
uniform acceleration, a particle in a harmonic oscillator potential, and
a system corresponding to an unstable equilibrium. We find, for specific 
choices of initial parameters, that as much as $90\%$ of the kinetic energy 
can be localized (at least conceptually) in the `front half' of such 
Gaussian wave packets, and we visualize these effects.

\end{abstract}

\maketitle

\section{\label{sec_intro} Introduction}

The study of time-dependent solutions of the Schr\"{o}dinger equation
for initial wave packets of Gaussian form has a long history, dating
back to the earliest days of the development of quantum theory.
Schr\"{o}dinger himself \cite{schrodinger} and others
\cite{kennard} - \cite{debroglie} used such solutions to discuss the
connections between the quantum and classical formulations of mechanics
and found explicit wave packet solutions to many of the standard
problems of classical mechanics, including the free-particle, case
of uniform acceleration, the harmonic oscillator, and the particle in
a uniform magnetic field. These examples first appeared in a number of 
textbooks \cite{kemble} - \cite{rojansky} less than a decade later.

In contrast, many treatments of introductory quantum mechanics  today focus 
almost exclusively on time-independent, energy eigenstate (or stationary 
state) solutions
of the Schr\"odinger equation, although a large number of texts do 
include the explicit construction of a time-dependent Gaussian wave packet
solution for the free-particle case. This easily obtained, analytic
result (seemingly first obtained by Darwin \cite{darwin}) 
illustrates, in a closed form solution, many of the notions 
central to wave packet spreading and the essentially classical 
time-development of the expectation value as 
$\langle x \rangle_t = \langle p\rangle_0 t/m + \langle
x \rangle_0$. In many modern software packages, these solutions are also 
visualized, helping
to illustrate the correlated  time-development of the real and imaginary
parts, and especially their phase relationships, often using a popular
color-coding scheme \cite{styer}, \cite{thaller}, \cite{belloni}.

An example of a static image representing such time-development is
shown in Fig.~1. One can easily note the non-symmetric pattern of 
the `wiggliness' apparent for $t>0$ in both the real (dotted) and 
imaginary (dashed) parts of 
$\psi(x,t)$, which, as we will argue below,  can be used as a qualitative 
measure of the local kinetic energy. The faster (higher momentum, and hence
`wigglier') components  are more obviously associated with the `front end' 
of the spreading wave packet, while the `back end' exhibits much
less rapid spatial variation, consistent with the slower (low momentum)
components trailing behind. 

In this note, we attempt to extend and amplify upon 
this type of intuitive  observation
and try to answer such questions as {\it Where is the kinetic energy
localized in a wave packet?} To this end, we begin with 
 the familiar observation that the expectation value of the 
kinetic energy operator in any time-dependent
position-space state, $\psi(x,t)$,  can be written, using a simple 
integration-by-parts (or IBP) argument, in the form
\begin{eqnarray}
\langle \hat{T}\rangle 
= \frac{1}{2m}\langle \hat{p}^2\rangle 
& = & -\frac{\hbar^2}{2m}
\integral dx \,\psi^*(x,t) \frac{\partial ^2 \psi(x,t)}{\partial x^2} \nonumber \\
& \stackrel{IBP}{=} & 
-\frac{\hbar^2}{2m}
\left(\psi^*(x,t) \frac{\partial \psi(x,t)}{\partial x}\right)^{+\infty}_{-\infty}      \\
& & \;\;\;\;\;\;\;+ \frac{\hbar^2}{2m}
\integral dx \,
\frac{\partial \psi^*(x,t)}{\partial x}
\, 
\frac{\partial \psi(x,t)}{\partial x} \nonumber 
\end{eqnarray}
or
\begin{equation}
\langle \hat{T} \rangle   =  \frac{\hbar^2}{2m}\integral dx 
\left|\frac{\partial \psi(x,t)}{\partial x}\right|^2 
\end{equation} 
where the `boundary terms' (involving $\psi$ and $\partial \psi/\partial x$ 
at $\pm \infty$) are assumed to vanish for any appropriately localized 
solution.

This global identification suggests that we define a local
{\it kinetic energy density}, ${\cal T}(x,t)$, via
\begin{equation}
{\cal T}(x,t) \equiv 
\frac{\hbar^2}{2m} 
\left|\frac{\partial \psi(x,t)}{\partial x} \right|^2
\;\;\; \;\;\;\;\;
\mbox{where}
\;\;\; \;\;\;\;\;
\langle \hat{T} \rangle_t =  \integral {\cal T}(x,t)\,dx 
\equiv T(t)
\, .
\label{kinetic_energy_distribution}
\end{equation}
This quantity, which is clearly locally real and positive-definite,
can then be used to quantify the distribution of kinetic energy, and
how it changes with time, for any time-dependent solution of the
Schr\"odinger equation. For Gaussian solutions, we will be able to
perform many of the desired integrals in closed form, leading to 
explicit analytic results for such related quantities as
\begin{equation}
T^{(+)}(t) \equiv  \int_{\langle x \rangle_t}^{+\infty} {\cal T}(x,t)\,dx 
\qquad
\mbox{and}
\qquad
T^{(-)}(t) \equiv  \int^{\langle x \rangle_t}_{-\infty} {\cal T}(x,t)\,dx 
\label{half_kinetic_energies}
\end{equation}
where the expectation value $\langle x \rangle_t$ serves to define the 
`center' of the wave
packet. These quantities can be respectively associated with the kinetic 
energy in the `right' (or `front', at least for packets moving generally
to the right) half and the `left' (or `back') half of the wave packet.

In what follows, we will present several explicit closed-form
examples of the
calculation of $T^{(\pm)}(t)$ for Gaussian wave packet solutions, 
starting in Sec.~\ref{sec_free_particle} with free-particle wave packets
for a standard Gaussian momentum distribution, while in 
Sec.~\ref{sec_uniform_acceleration} we illustrate 
similar results for Gaussian solutions to the problem of a particle 
undergoing uniform acceleration, working initially in momentum-space.
In Sec.~IV  we use standard propagator techniques to examine Gaussian wave 
packet solutions for the harmonic oscillator problem in this context,
while in Sec.~V we extend these results to the case of an
`inverted oscillator', corresponding to a  particle in unstable equilibrium.

\section{\label{sec_free_particle} Free-particle Gaussian wave packets}

The time-dependent Schr\"odinger equation for the one-dimensional free
particle case can be written, and easily solved, in either position- or
momentum-space in the equivalent forms
\begin{equation}
- \frac{\hbar^2}{2m} \frac{\partial ^2\psi(x,t)}{\partial x^2}
= i \hbar \frac{\partial \psi(x,t)}{\partial t}
\qquad
\quad
\mbox{or}
\qquad
\quad
\frac{p^2}{2m} \phi(p,t) = i \hbar \frac{\partial \phi(p,t)}{\partial t}
\, .
\end{equation}
Since we will find the momentum-space approach more useful in 
Sec.~\ref{sec_uniform_acceleration}
for the case of uniform acceleration, we will also use that approach here
and write
\begin{equation}
\phi(p,t) = \phi_{0}(p) e^{-ip^2t/2m\hbar}
\end{equation}
where $\phi(p,0) = \phi_{0}(p)$ is the initial momentum distribution.
Using this very general form, and the appropriate operator form of
$\hat{x} = i\hbar (\partial/\partial p)$, we recall that
\begin{eqnarray}
\langle \hat{x} \rangle_t & = & \int_{-\infty}^{+\infty}
\phi^*(p,t) \, \left(i\hbar \frac{\partial}{\partial p}\right)
\phi(p,t) \,dp \nonumber \\
& = &
\int_{-\infty}^{+\infty} \phi_{0}^*(p) \hat{x} \phi_{0}(p)\,dp
+ \frac{t}{m} \int_{-\infty}^{+\infty} p |\phi_{0}(p)|^2\,dp \\
& = & \langle \hat{x} \rangle_0 + \frac{\langle p\rangle_0 t}{m}
\, .
\nonumber
\end{eqnarray}
The position-space solution can be written, of course,
using the Fourier transform as
\begin{equation}
\psi(x,t) = \pre \integral e^{ipx/\hbar} \, \phi(p,t)\,dp
\, .
\label{fourier_transform}
\end{equation}
The standard initial Gaussian momentum-space distribution,
 which gives arbitrary initial momentum ($p_0$) 
and position ($x_0$) values,
 can be written in the form
\begin{equation}
\phi(p,0) = \phi_0(p) = 
\sqrt{\frac{\alpha}{\sqrt{\pi}}}
\, e^{-\alpha^2(p-p_0)^2/2}
\, e^{-ipx_0/\hbar}
\label{initial_gaussian}
\end{equation}
which gives
\begin{equation}
\langle p \rangle_{t} = p_0
\, , 
\qquad
\quad
\langle p^2 \rangle_{t} = p_0^2 + \frac{1}{2\alpha^2}
\, ,
\qquad
\mbox{and}
\qquad
\Delta p_t = \Delta p_0 = \frac{1}{\alpha \sqrt{2}}
\, .
\end{equation}
The explicit form of the position-space wave function is given by
the Gaussian integral
\begin{equation}
\psi(x,t) = \pre \sqrt{\frac{\alpha}{\sqrt{\pi}}}
\integral\, e^{ip(x-x_0)/\hbar}\,
e^{-\alpha^2 (p-p_0)^2/2}\,
e^{-ip^2t/2m\hbar}\,dp
\end{equation}
which can be evaluated in closed form (using the change of variables
$q \equiv p-p_0$ and standard integrals) to obtain
\begin{equation}
\psi(x,t) = \frac{1}{\sqrt{\sqrt{\pi} \alpha \hbar (1+it/t_0)}}
\,
e^{ip_0(x-x_0)/\hbar}
\, e^{-ip_0^2t/2m\hbar}
\,
e^{-(x-x_0-p_{0}t/m)^2/2(\alpha \hbar)^2(1+it/t_0)}
\label{free_particle_position_solution}
\end{equation}
where $t_0 \equiv m\hbar \alpha^2$. (This result is sometimes 
attributed to Darwin \cite{darwin}.)
The corresponding probability density is easily shown to be
\begin{equation}
P(x,t) = |\psi(x,t)|^2 = \frac{1}{\sqrt{\pi} \beta_t}
e^{-(x-x(t))^2/\beta_t^2}
\end{equation}
where
\begin{equation}
x(t) \equiv x_0 + p_0t/m
\qquad
\mbox{and}
\qquad
\beta_t \equiv \alpha \hbar \sqrt{1+ t^2/t_0^2}
\end{equation}
and the time-dependent expectation values of position are 
\begin{equation}
\langle x \rangle_t = x(t) = x_0 + p_0t/m 
\, ,
\qquad
\quad
\langle x^2 \rangle_t = (x(t))^2 + \frac{\beta^2_t}{2}
\, ,
\qquad
\mbox{and}
\quad
\Delta x_t = \frac{\beta_t}{\sqrt{2}}
\, ,
\end{equation}
all of which are familiar results.

Turning now to the kinetic energy distribution defined in 
Eqn.~(\ref{kinetic_energy_distribution}), we find that the required
spatial derivative is given by
\begin{equation}
\frac{\partial \psi(x,t)}{\partial x}
= \left(\frac{ip_0}{\hbar} - \frac{ (x-x(t))}{(\alpha \hbar)^2 (1+it/t_0)}
\right) \psi(x,t)
\, .
\end{equation}
The kinetic energy density can therefore be written in the form
\begin{equation}
{\cal T}(x,t) = \frac{1}{2m}
\left( p_0^2 + \left[\frac{2(x-x(t)) p_0}{\alpha^2\hbar}\right]
\left[\frac{t/t_0}{(1+t^2/t_0^2)}\right]
+ \frac{(x-x(t))^2}{(\alpha^2 \hbar)^2 (1+t^2/t_0^2)}\right)
|\psi(x,t)|^2
\, . 
\label{gaussian_case}
\end{equation}
The expectation value of the kinetic energy is correctly given by
\begin{equation}
T(t) = \integral\, {\cal T}(x,t)\,dx = \frac{1}{2m} 
\left(p_0^2 + \frac{1}{2\alpha^2}\right)
\end{equation}
and receives non-zero contributions from the first and last terms in brackets
in Eqn.~(\ref{gaussian_case}),  since the middle term vanishes (when
integrated over all space) for symmetry reasons. On the other hand,
the individual values of $T^{(\pm)}(t)$ in 
Eqn.~(\ref{half_kinetic_energies}) can also be evaluated giving
\begin{equation}
T^{(\pm)}(t) 
= \frac{1}{2m}
\left(\frac{1}{2}\right)
\left( 
p_0^2 
\pm 
\left(\frac{2p_0}{\alpha \sqrt{\pi}} \right) \frac{t/t_0}{\sqrt{1+t^2/t_0^2}} 
+ \frac{1}{2\alpha^2}
\right)
\label{left_and_right_kinetic_energies}
\end{equation}
both of which are easily seen to be positive definite, as they must,
due to the non-negativity of ${\cal T}(x,t)$. The time-dependent fractions
of the total kinetic energy contained in the $(+)/(-)$ (right/left) halves 
of this archetypical wave packet are given by
\begin{equation}
R^{(\pm)}(t) \equiv \frac{T^{(\pm)}(t)}{T^{(+)}(t) + T^{(-)}(t)}
= \frac{1}{2} \pm 
 \left(\frac{2}{\sqrt{\pi}}\right)
\left( \frac{(p_0\alpha)}{(2(p_0\alpha)^2+1)}\right) 
\frac{t/t_0}{\sqrt{1+t^2/t_0^2}}
\label{define_r_function}
\end{equation}
which will clearly increase/decrease monotonically as 
$t/t_0 \rightarrow \infty$.
The limiting values are then
\begin{equation}
R^{(\pm)}(t/t_0\rightarrow \infty) = \frac{1}{2} \pm 
\left(\frac{2}{\sqrt{\pi}}\right) \left(\frac{(p_0\alpha)}{2(p_0\alpha)^2+1}
\right)
\end{equation}
which have the  extremal values
\begin{equation}
R^{(+)}_{max}(t/t_0 \rightarrow \infty) 
 =  \frac{1}{2} + \frac{1}{\sqrt{2\pi}} \approx 0.9
\qquad
\mbox{and}
\qquad
R^{(-)}_{min}(t/t_0 \rightarrow \infty) 
 =  \frac{1}{2} - \frac{1}{\sqrt{2\pi}} \approx 0.1
\label{maximum_minimum_values}
\end{equation}
which occur when
\begin{equation}
p_0 \alpha = \frac{1}{\sqrt{2}}
\qquad
\mbox{or}
\qquad
p_0 = \frac{1}{\alpha \sqrt{2}} = \Delta p_0
\label{extreme_value}
\, . 
\end{equation}
Thus, as much as $90\%$ of the total kinetic energy of this Gaussian
wave packet solution can be in the `front half' of the wave at long times
(i.e. those for which $t >> t_0$.)

To illustrate this effect, we plot in the left column of Fig.~2,  
the modulus ($|\psi(x,t)|$, solid),
and real (dotted) and imaginary (dashed) parts of a typical solution
corresponding to $p_0 = 0$ (top), 
$p_0 = \Delta p_0$ (the extremal value, middle), and
$p_0 = 4\Delta p_0$ (bottom) for long times ($t = 10t_0$).
The values are plotted in terms of the variable
$x(t) = x_0 + p_0t/m$ for easier comparison. For the $p_0=0$ case, it is
clear that the larger momentum components (in magnitude) spread faster,
but uniformly, in the opposite $+x$ and $-x$ directions, giving 
equal `wiggliness' on each 
side, while for large values of $p_0$, the total kinetic energy is clearly
increased (many more `wiggles' everywhere),  
but the amount on each side of the
expectation value peak at $\langle x \rangle_t$ is  roughly the same.
For the extremal value of $p_0 = \Delta p_0$, there is the clearest
distinction between the `front' and `back' halves,  as the magnitudes
of the momentum components in the `back' half are at a minimum, resulting
in the greatest separation between the kinetic energy in the two halves.

We can also compare the distribution of probability, described as usual
by $P(x,t) \equiv |\psi(x,t)|^2$, to how the kinetic energy is localized.
The kinetic energy density, ${\cal T}(x,t)$, can be scaled to the total
(and possibly time-dependent) value of $T(t)$ via 
\begin{equation}
S(x,t) \equiv \frac{{\cal T}(x,t)}{T(t)}
\qquad
\mbox{which satisfies}
\qquad
\integral\, S(x,t)\,dx = 1
\label{scaled_kinetic_energy_density}
\end{equation}
and so is normalized in the same way as the probability density.
We plot both $P(x,t)$ (solid curve) and $S(x,t)$ (dot-dashed curve) 
in the right hand column of Fig.~2 for the same three cases considered above
and we note how the shape of $S(x,t)$ is correlated with the obvious 
`wiggliness'
shown in the real and imaginary parts of $\psi(x,t)$.

\section{\label{sec_uniform_acceleration} Uniform acceleration}

The problem of a particle under the influence of a constant
force is a staple in  classical mechanics,  and was considered early
in the history of quantum mechanics \cite{kemble}, but is less often 
discussed in introductory treatments of the subject,  especially in terms 
of time-dependent solutions. For that reason, we briefly review the most
straightforward momentum-space approach to this problem. In this case,
where the potential is given by $V(x) = -Fx$, we can write the
time-dependent Schr\"odinger equation in momentum-space as
\begin{equation}
\frac{p^2}{2m}\phi(p,t) - 
F\cdot \left[i\hbar \frac{\partial}{\partial p}\right] \phi(p,t)
= i\hbar \frac{\partial \phi(p,t)}{\partial t}
\label{pspace}
\end{equation}
or 
\begin{equation}
i\hbar\left(F\frac{\partial \phi(p,t)}{\partial p} 
+ \frac{\partial \phi(p,t)}{\partial t}\right) = \frac{p^2}{2m}\phi(p,t)
\, . 
\label{newp}
\end{equation}
We note that the simple combination of derivatives guarantees that a
function of the form $\Phi(p-Ft)$ will make the left-hand side vanish,  so
we assume a solution of the form $\phi(p,t) = \Phi(p-Ft)\tilde{\phi}(p)$,
with $\Phi(p)$ arbitrary and $\tilde{\phi}(p)$ to be determined.
Using this form, Eqn.~(\ref{newp}) reduces to
\begin{equation}
\frac{\partial \tilde{\phi}(p)}{\partial p} = 
-\frac{i p^2}{2m\hbar F}\tilde{\phi}(p)
\end{equation}
with the solution
\begin{equation}
\tilde{\phi}(p) = e^{-ip^3/6mF\hbar}
\, . 
\end{equation}
We can then write the general solution as
\begin{equation}
\phi(p,t) = \Phi(p\!-\!Ft)e^{-ip^3/6mF\hbar}
\end{equation}
or, using the arbitrariness of $\Phi(p)$, as
\begin{equation}
\phi(p,t) = \phi_0(p\!-\!Ft) e^{i((p-Ft)^3-p^3)/6mF\hbar )}
\label{pacc}
\end{equation}
where now $\phi_0(p)$ is some initial momentum distribution since
$\phi(p,0) = \phi_0(p)$. Note that because the $p^3$ terms cancel
in the exponential, we will be able to explicitly integrate
Gaussian type initial momentum-space waveforms.

For any general initial $\phi_{0}(p)$ we now have the time-dependent
expectation values
\begin{eqnarray}
\langle p \rangle_t &= & \langle p \rangle_0 + Ft \\
\langle p^2 \rangle_t &= & \langle p^2 \rangle_0 + 
2\langle p \rangle_0 Ft +(Ft)^2 \\
\langle \hat{x} \rangle_t & = &  \langle \hat{x} \rangle_0
+ \frac{\langle p\rangle_0 t}{m} + \frac{Ft^2}{2m} 
\end{eqnarray}
which also give the expectation value
\begin{equation}
\left \langle \frac{p^2}{2m} + V(x) \right\rangle = 
\frac{\langle p^2 \rangle_0}{2m}
- F\langle \hat{x} \rangle_0
\end{equation}
which, in turn, also  agrees with a similar calculation of 
$\langle \hat{E} \rangle_t$
using $\hat{E} = i\hbar (\partial/\partial t)$,  all of which are consistent
with a particle undergoing uniform acceleration.

Using the standard initial Gaussian momentum-space wavefunction in 
Eqn.~(\ref{initial_gaussian}) as the $\phi_0(p-Ft)$ in Eqn.~(\ref{pacc}), 
we can evaluate the position-space solution
using Eqn.~(\ref{fourier_transform}) to obtain
\begin{eqnarray}
\psi(x,t) & = & \left[e^{iFt(x_0-Ft^2/6m)/\hbar}\, e^{i(p_0+Ft)(x-x_0 - p_0t/2m)/\hbar}\right] \left(
\frac{1}{\sqrt{\sqrt{\pi}\alpha \hbar (1+it/t_0)}}
\right) \nonumber \\ 
& & \,\,\,
\times
\, e^{-(x-(x_0+p_0t/m+Ft^2/2m))^2/2(\alpha \hbar)^2(1+it/t_0)}
\, . 
\label{accelerating_solution}
\end{eqnarray}
The corresponding probability density is given by
\begin{equation}
P(x,t) = \frac{1}{\sqrt{\pi} \beta_t}
e^{-(x-\tilde{x}(t))^2/\beta_t^2}
\end{equation}
where 
\begin{equation}
\tilde{x}(t) \equiv x_0 + \frac{p_0t}{m} + \frac{Ft^2}{2m}
\end{equation}
and 
\begin{equation}
\qquad
\langle x \rangle_t = \tilde{x}(t)
\, ,
\qquad
\langle x^2 \rangle_t = (\tilde{x}(t))^2 + \frac{\beta_t^2}{2}
\, ,
\quad
\mbox{and}
\qquad
\Delta x_t = \frac{\beta_t}{\sqrt{2}} 
\end{equation}
so that this accelerating wave packet spreads in the same manner as the
free-particle Gaussian example. 
The calculation of the kinetic energy density proceeds exactly 
as in Sec.~\ref{sec_free_particle}, with
\begin{eqnarray}
{\cal T}(x,t) & = &  \frac{1}{2m}
\left( (p_0+Ft)^2
+ \frac{2(p_0+Ft)(x-\tilde{x}(t))}{\hbar \alpha^2}
\frac{t/t_0}{(1+t^2/t_0^2)} \right. \nonumber \\
& &
\,\,\,\,\,\,\,\,\,\,\,\,\,\,\,\,\,\,\,\,\,\,\,\,\,\,\,\,\,
\left.
+ \frac{(x-\tilde{x}(t))^2}{(\alpha^2\hbar)^2 (1+t^2/t_0^2)}
\right)
|\psi(x,t)|^2
\end{eqnarray}
The corresponding $(+)$ and $(-)$ kinetic energies are then derived
in the same way as before and are given by 
\begin{equation}
T^{(\pm)}(t)
= \frac{1}{2m} \left(\frac{1}{2}\right)
\left( (p_0+Ft)^2 
\pm 
\left( \frac{2(p_0+Ft)}{\alpha\sqrt{\pi}}\right)
\left(\frac{t/t_0}{\sqrt{1+t^2/t_0^2}}\right)
+
\frac{1}{2\alpha^2}
\right)
\end{equation}
which is the same result as in Eqn.~(\ref{left_and_right_kinetic_energies}),
with $p_0 \rightarrow p_0 +Ft$. This similarity in form implies that the
maximum (minimum) values of $T^{(\pm)}(t)$ are once again given by
Eqn.~(\ref{maximum_minimum_values}) (provided that $t/t_0 >> 1$) 
which now  occurs when 
$|p_0 +Ft| = \Delta p_0$. In this more dynamic situation, if $p_0$
and $F$ have different signs (so that the motion includes one `back'
and one `forth' component), this situation can occur twice during
a single trajectory, with the roles of $R^{(+)}(t)$ and
$R^{(-)}(t)$ changing between the `back' and the `forth' traversal.

\section{\label{sec_sho} Simple harmonic oscillator wave packets}

The third model system which is easily shown to exhibit time-dependent
Gaussian wave packet solutions
 is also the most frequently studied of all classical
and quantum mechanical examples, namely the simple harmonic oscillator,
defined by the potential energy $V(x) = m \omega^2 x^2/2$. In this case,
the initial value problem is perhaps most easily solved, especially for
Gaussian wave packets, by propagator techniques \cite{feynman_1} --
\cite{holstein}. In this approach, one writes
\begin{equation}
\psi(x,t) = \int_{-\infty}^{+\infty} \,dx'\, \psi(x',0)\, K(x,x';t,0)
\label{propagator}
\end{equation}
where the propagator can be derived in a variety of ways \cite{holstein}
and can be written in the form
\begin{equation}
K(x,x';t,0) = \sqrt{\frac{m \omega}{2\pi i \hbar \sin(\omega t)}}
\exp
\left[
\frac{im \omega}{2\hbar \sin(\omega t)} ((x^2 + (x')^2)\cos(\omega t)
- 2x x')
\right]
\, .
\end{equation}
For the initial state wavefunction we will use position-space
version of Eqn.~(\ref{initial_gaussian}), but for notational and
visualization simplicity, we will specialize to the case of $x_0 = 0$, 
namely
\begin{equation}
\psi(x',0) = \frac{1}{\sqrt{\beta \sqrt{\pi}}} \, e^{ip_0x'/\hbar}\,
e^{-(x')^2/2\beta^2}
\label{initial_position_gaussian}
\end{equation}
where $\beta = \alpha \hbar$ and $\Delta x_0 = \beta/\sqrt{2}$. 
In this state,  the expectation value of the energy is
\begin{equation}
\langle \hat{E} \rangle_t = \langle \hat{E} \rangle_0 =
\frac{1}{2m} \left(p_0^2 + \frac{\hbar^2}{2\beta^2}\right)
+ \frac{m\omega^2\beta^2}{4}
\label{total_sho_energy}
\end{equation}
The integral in Eqn.~(\ref{propagator}) can be done in closed form for
the initial Gaussian in Eqn.~(\ref{initial_position_gaussian}) with the
result
\begin{equation}
\psi(x,t) =
\exp
\left[
\frac{im\omega x^2 \cos(\omega t)}{2\hbar \sin(\omega t)}
\right]
\frac{1}{\sqrt{A(t) \sqrt{\pi}}}
\exp
\left[ 
-\frac{i m \omega \beta}{2\hbar \sin(\omega t)}
\frac{(x-x_s(t))^2}{A(t)}
\right]
\label{position_space_sho_solution}
\end{equation}
where
\begin{equation}
A(t) \equiv \beta \cos(\omega t) + i \left(\frac{\hbar}{m \omega \beta}
\right) \sin(\omega t)
\qquad
\mbox{and}
\qquad
x_s(t) \equiv \frac{p_0 \sin(\omega t)}{m \omega}
\, .
\end{equation}
(We note that in the force-free limit, when  $\omega \rightarrow 0$,
one can show that this solution reduces to the free-particle 
form  in Eqn.~(\ref{free_particle_position_solution}), with $x_0 = 0$,
as expected.)
The time-dependent probability density is given by
\begin{equation}
P(x,t) = |\psi(x,t)|^2
= \frac{1}{\sqrt{\pi}|A(t)|} e^{-(x-x_s(t))^2/|A(t)|^2}
\end{equation}
where
\begin{equation}
|A(t)| = \sqrt{\beta^2 \cos^2(\omega t) + (\hbar/m\omega \beta)^2
\sin^2(\omega t)}
\end{equation}
and
\begin{equation}
\langle x\rangle_t = x_s(t)
\qquad
\mbox{and}
\qquad
\Delta x_t = \frac{|A(t)|}{\sqrt{2}}
\, .
\end{equation}
So, while the expectation value of the wave packet oscillates in a
way which mimics the classical result, the time-dependent spatial
width of the packet changes in time, consistent with very
general expectations for the oscillator case \cite{gottfried},
\cite{styer_2}.
This more general time-dependent Gaussian solution is described
in several textbooks \cite{rojansky}, \cite{saxon} and
special cases of it are often rediscovered \cite{pulsate},
\cite{pulsate_2}.

We note that in the special case of the minimum uncertainty wavepacket
where 
\begin{equation}
\beta^2 = \left(\frac{\hbar}{m\omega \beta}\right)^2
\qquad
\mbox{or}
\qquad
\beta = \sqrt{\frac{\hbar}{m\omega}} \equiv \beta_0
\end{equation}
the time-dependent width of the Gaussian simplifies to
\begin{equation}
\Delta x_t = \frac{\beta_0}{\sqrt{2}}
= \Delta x_0
\end{equation}
so the wave packet oscillates with no change in shape. This is
the special result seen more standardly \cite{schiff} -- \cite{roy}
in textbooks and pedagogical articles, and is similar to the
coherent-state like solution discussed by
Schr\"{o}dinger \cite{schrodinger} in a famous paper.

The time-dependent
expectation value of the potential energy is easily found to be
\begin{equation}
\langle V(x)\rangle_t
= \frac{1}{2} m\omega^2 \langle x^2 \rangle_t
= 
\frac{1}{2m}\left( p_0^2 + \frac{\hbar^2}{2\beta^2}\right) \sin^2(\omega t)
+ \frac{m \omega^2 \beta^2 \cos^2(\omega t)}{4}
\, . 
\end{equation}
The time-dependent (total) kinetic energy then follows directly from
this equation combined with  Eqn.~(\ref{total_sho_energy}) and is given by
\begin{equation}
T(t) = \frac{1}{2m}\langle p^2 \rangle_t
= \frac{1}{2m} \left( p_0^2 + \frac{\hbar^2}{2\beta^2}\right) \cos^2(\omega t)
+ 
\frac{m \omega^2 \beta^2 \sin^2(\omega t)}{4}
\, .
\end{equation}
In order to determine the {\it distribution} of kinetic energy, however, 
we evaluate ${\cal T}(x,t)$ using
\begin{equation}
\frac{\partial \psi(x,t)}{\partial x}
= 
\left(
\frac{im\omega}{\hbar \sin(\omega t)}
\right)
\left( 
x \cos(\omega t) - \frac{\beta (x-x_s(t))}{A(t)}
\right)
\psi(x,t)
\end{equation}
and we find that
\begin{equation}
T^{(\pm)}(t) 
= \frac{T(t)}{2}
\pm 
\frac{p_0 \omega \sin(\omega t) \cos^2(\omega t)}{2\sqrt{\pi} |A(t)|}
\left[
\frac{\beta_0^4}{\beta^2} - \beta^2
\right]
\label{sho_fraction}
\, .
\end{equation}
The asymmetry ($T^{(+)}(t) \neq T^{(-)}(t)$) in the kinetic energy
distribution vanishes at half-integral multiples of the classical period
$\tau \equiv 2\pi/\omega$, namely when $t = n\tau/2$, but also at the
classical turning points, {\it i.e.}, at odd multiples of $\tau/4$. 
There is also no asymmetry in the case when $p_0 = 0$ and the wave packet 
expands and contracts uniformly in both the $+x$ and $-x$ directions.
Finally, there is no asymmetry in the special case of the `fixed width' 
Gaussian, when $\beta = \beta_0$, and this property can perhaps help
explain some of the special features of that minimum-uncertainty state.

To better visualize the general result in Eqn.~(\ref{sho_fraction}), 
we plot in Figs.~3 and 4 representations
of both the wave function (modulus, real, and imaginary parts) as well
as the probability ($P(x,t)$) and (scaled) kinetic energy 
($S(x,t)$) distributions over the
first quarter period. We note that for wave packets initially moving
to the right ($p_0 >0$) as shown here, narrow packets, i.e., ones
with $\beta < \beta_0$, typically have  more kinetic energy in the 
`front' half of the packet (Fig.~3, middle panels), while initially
wider packets have the opposite behavior (Fig.~4) consistent with
Eqn.~(\ref{sho_fraction}).

While the time-dependence  of $R^{(\pm)}(t)$ 
(defined in Eqn.~(\ref{define_r_function})) is more
varied than for the simpler, non-periodic, cases considered so far, 
as a specific example, we can examine the distribution of kinetic energy at 
times such
that 
\begin{equation}
\cos(\omega T) = \sin(\omega T) = \frac{1}{\sqrt{2}}
\qquad
\mbox{for instance, when}
\qquad
T = \frac{\tau}{8}
\, . 
\end{equation}
In that case we find
\begin{equation}
R^{(\pm)}(T) =
\frac{1}{2}
\pm
\left(
\frac{2}{\sqrt{\pi}}
\right)
\left(\frac{ (p_0/m\omega)}{(2(p_0/m\omega)^2 + (\beta_0^4/\beta^2
+ \beta^2)}\right)
\left(
\frac{(\beta_0^4/\beta^2) - \beta^2)}
{\sqrt{(\beta_0^4/\beta^2) + \beta^2)}}
\right)
\end{equation}
This has extremal values of
\begin{equation}
R^{(\pm)}(T) =
\frac{1}{2}
\pm
\frac{1}{\sqrt{2\pi}}
\left[
\frac{\beta_0^4/\beta^2 - \beta^2}
     {\beta_0^4/\beta^2 + \beta^2}
\right]
\end{equation}
when
\begin{equation}
p_0 =  \sqrt{\frac{(\beta m\omega)^2}{2} + \frac{\hbar^2}{2\beta^2}}
\label{sho_extreme_value}
\end{equation}
which are obvious generalizations of Eqns.~(\ref{maximum_minimum_values})
and (\ref{extreme_value}).

\section{\label{sec_unstable} `Inverted' oscillator wave packets for 
unstable equilibrium}

The final case we present for which time-dependent Gaussian wave packet 
solutions are easily obtained is a generalization of the harmonic oscillator
which has been described as the `inverted oscillator' \cite{nardone}
or the `unstable particle' \cite{robinett_2}, corresponding to
the classical motion of a particle at the top of potential hill, given
by 
\begin{equation}
\tilde{V}(x) \equiv -\frac{1}{2} m\tilde{\omega}^2 x^2
\, . 
\end{equation}
Many of the results obtained using propagator techniques can be easily
carried over to this problem with the simple identifications
\begin{equation}
\omega \rightarrow i \tilde{\omega}
\, ,
\qquad
\sin(\omega t) \rightarrow i \sinh(\tilde{\omega} t)
\, ,
\qquad
\mbox{and}
\qquad
\cos(\omega t) \rightarrow \cosh(\tilde{\omega} t)
\, . 
\end{equation}
For example, the position-space probability density corresponding
to the initial state in Eqn.~(\ref{initial_position_gaussian}) 
is given by
\begin{equation}
P(x,t) = |\psi(x,t)|^2
= \frac{1}{\sqrt{\pi}|B(t)|} \exp^{-(x-\tilde{x}_s(t))^2/|B(t)|^2}
\end{equation}
where
\begin{equation}
|B(t)| = \sqrt{\beta^2 \cosh^2(\tilde{\omega} t) + 
(\hbar/m\tilde{\omega} \beta)^2
\sinh^2(\tilde{\omega} t)}
\end{equation}
and
\begin{equation}
\langle x\rangle_t = \tilde{x}_s(t) = 
\frac{p_0 \sinh(\tilde{\omega}t)}{m\tilde{\omega}}
\qquad
\mbox{and}
\qquad
\Delta x_t = \frac{|B(t)|}{\sqrt{2}}
\end{equation}
and the probability density exhibits a `runaway' (exponential)
behavior. 
In this case, the long time behavior of the kinetic energy fractions
is dictated by the limits
\begin{equation}
\cosh(\tilde{\omega}t), \,\sinh(\tilde{\omega} t)
\longrightarrow 
e^{\tilde{\omega}t}/2
\end{equation}
which gives
\begin{equation}
R^{(\pm)}(t \rightarrow \infty)
= \frac{1}{2}
\pm
\left(\frac{2}{\sqrt{\pi}}\right)
\left[
\frac{(p_0/m\tilde{\omega})}
{ 2(p_0/m\tilde{\omega})^2 + (\beta_0^4/\beta^2 + \beta^2)}
\right]
\sqrt{\beta_0^4/\beta^2 + \beta^2}
\end{equation}
with the same extremal values as in Eqn.~(\ref{maximum_minimum_values}),
when $p_0$ satisfies the equivalent of Eqn.~(\ref{sho_extreme_value})
with $\omega \rightarrow \tilde{\omega}$.

\section{\label{sec_conclusions} Conclusions and discussion}

In this note we have provided a detailed analysis of Gaussian wave packet
solutions of the Schr\"odinger equation in several model systems, helping to
elucidate  some of the qualitative aspects of wave packet time-development
and spreading  often seen in standard visualizations, 
and their relationship to the distribution of
kinetic energy. We have focused on closed-form analytic results,  
all of which have been obtainable due to the special nature of the integrals 
which arise for
Gaussian wave packets. An obvious and interesting extension would be to
extend these results to free particle wave packets for arbitrary 
non-Gaussian initial momentum distributions, performing the required 
integrals in Eqn.~(\ref{fourier_transform}) numerically to see just
how general the results in Eqns.~(\ref{maximum_minimum_values})
and (\ref{extreme_value}) are. For example, are there other initial
wave packets and initial conditions for which the kinetic energy can
be even more localized that the $90\%/10\%$ fraction seen here for
Gaussians? Another straightforward generalization is to consider 
two-dimensional extensions of all four systems discussed here 
(then visualized as functions of both $x$ and $y$) where closed form
expressions are also possible, as well as the related case of the
charged particle (confined to a plane) subject to a uniform magnetic
field.

Other studies of dynamical wave packet propagation have
discussed aspects of the time-development of $\psi(x,t)$ which
arise from effects related to the differing behavior of various
momentum-components. Examples have included discussions of 
the average speed of wave packet components
which are transmitted or reflected from a rectangular barrier
\cite{bramhall} (due to the energy dependence of probabilities of 
transmission and reflection) as well as the time-dependent shape of 
$|\phi(p,t)|^2$ for a wave packet hitting an infinite wall \cite{doncheski}
or similar `bouncing' systems \cite{bouncing_bob}.
The consideration of the distribution of kinetic energy distribution
in many other such time-dependent wave packet solutions might prove useful
in understanding their behavior. One such example certainly would be
Gaussian-like wave packet solutions in the infinite square well, where 
additional interesting features, such as exact wave packet revivals, are 
present.

Gaussian solutions such as those considered here can find use as examples 
of model systems in discussing other theoretical constructs, such as the 
Wigner quasi-probability distribution 
\cite{wigner} - \cite{belloni_doncheski_robinett}.
In that case, one defines $P_{W}(x,p;t)$ via
\begin{eqnarray}
P_{W}(x,p;t)
 & \equiv &
\frac{1}{\pi \hbar}
\int_{-\infty}^{+\infty}
\psi^{*}(x+y,t)\,\psi(x-y,t)\,e^{2ipy/\hbar}\,dy \\
& = & 
\frac{1}{\pi \hbar}
\int_{-\infty}^{+\infty}
\phi^*(p+q,t)\, \phi(p-q,t)\, e^{-2ixq/\hbar}\,dq
\end{eqnarray}
which can then be evaluated explicitly 
in closed form, in either position- or 
momentum-space, for all of the solutions discussed above. In an accompanying 
paper \cite{bassett} we study the same four examples considered here, 
in yet a different context.

 \newpage

\begin{flushleft}
{\Large {\bf 
Figure Captions}}
\end{flushleft}
\vskip 0.5cm
 
\begin{itemize}
\item[Fig.\thinspace 1.]  Time-development of a free-particle Gaussian wave 
packet, illustrating the modulus ($|\psi(x,t)|$, solid), 
the real ($Re(\psi(x,t))$, dotted)
and imaginary ($Im(\psi(x,t))$, dashed) parts,  as functions of time.
\item[Fig.\thinspace 2.] The figures in the left column show the modulus 
(solid),
and the real (dotted) and imaginary (dashed) parts of Gaussian
wave packets at long times ($t = 10t_0$) plotted versus $x_0 + p_0 t/m$,
illustrating the large fraction of kinetic energy (note the pattern
of `wiggliness') in the front of the wave for the $p_0 = \Delta p_0$
case, as in Eqn.~(\ref{maximum_minimum_values}). In the right column, 
we plot the probability density ($|\psi(x,t)|^2$, solid) and the (scaled)
kinetic energy density ($S(x,t) = {\cal T}(x,t)/T(t)$, dashed)
from Eqn.~(\ref{scaled_kinetic_energy_density}).
\item[Fig.\thinspace 3.] The figures in the top row show the modulus (solid),
and the real (dotted) and imaginary (dashed) parts of Gaussian wave packet
solutions of the harmonic oscillator corresponding to $x_0 = 0$ and
$p_0 > 0$ with $\beta = \beta_0/2 < \beta_0$. Times over the first quarter
period are shown. In the bottom row, we plot the probability density ($|\psi(x,t)|^2$, solid) and the (scaled) kinetic energy density 
($S(x,t) = {\cal T}(x,t)/T(t)$, dashed) 
from Eqn.~(\ref{scaled_kinetic_energy_density}) at the corresponding times.
\item[Fig.\thinspace 4.]  Same as Fig.~3, but for an initial Gaussian
wave packet solution of the harmonic oscillator with 
$\beta = 2\beta_0 > \beta_0$.
\end{itemize}
 
\newpage

\noindent
\hfill
\begin{figure}[hbt]
\, \hfill \,
\begin{minipage}{0.7\linewidth}
\epsfig{file=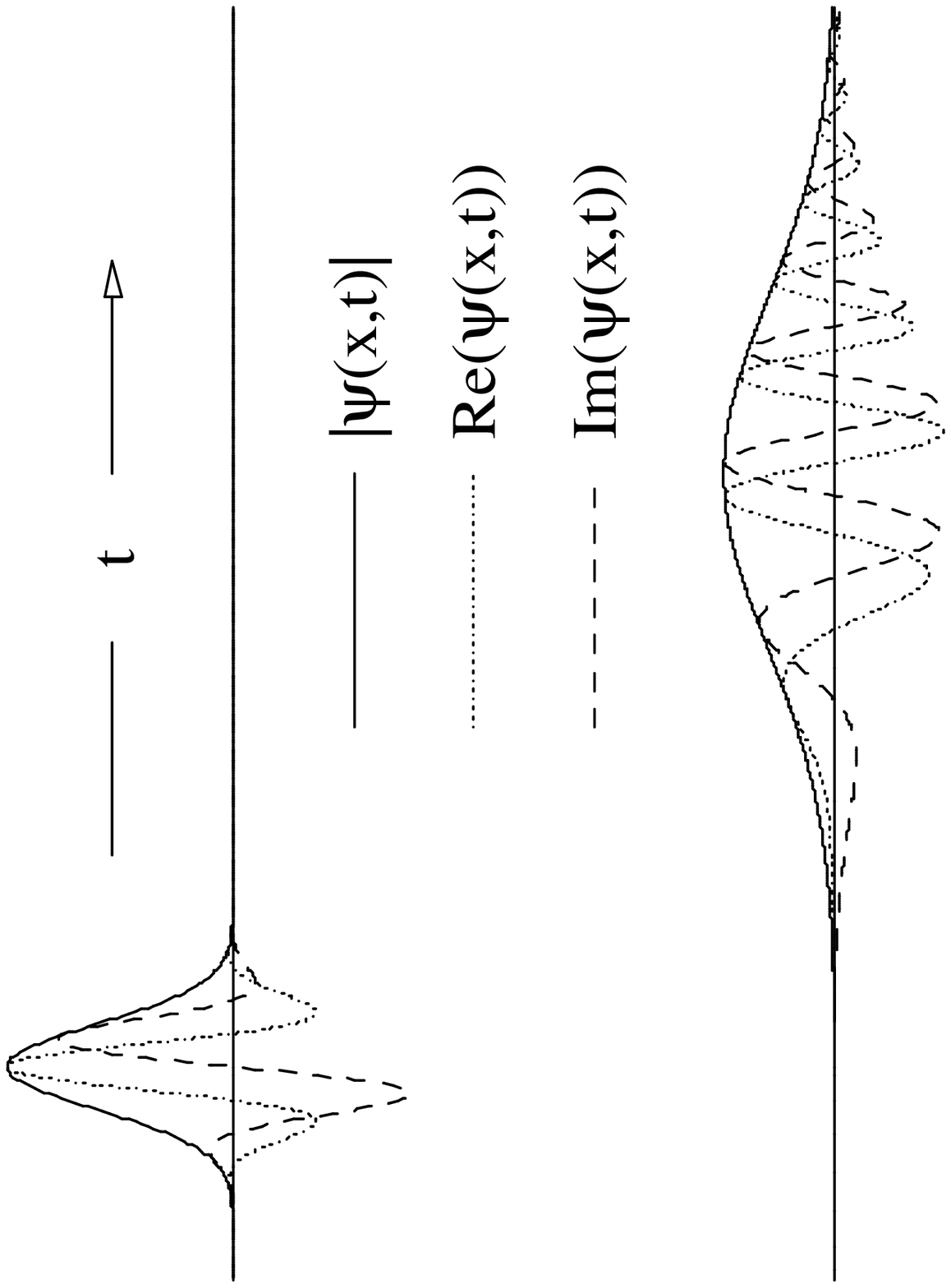,width=\linewidth}
\caption{}
\end{minipage}
\, \hfill \,
\end{figure}
\hfill

\newpage

\noindent
\hfill
\begin{figure}[hbt]
\, \hfill \,
\begin{minipage}{0.7\linewidth}
\epsfig{file=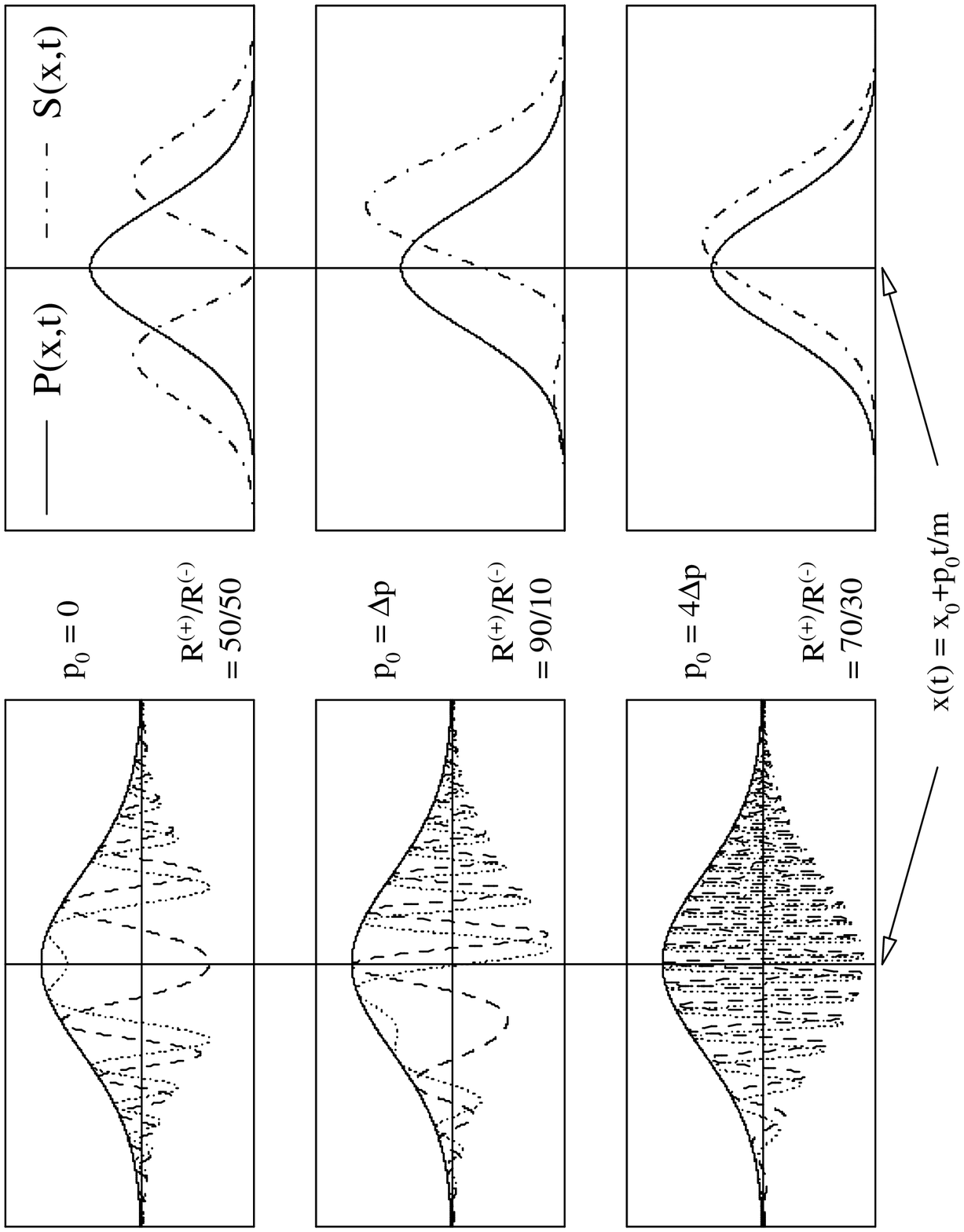,width=\linewidth}
\caption{}
\end{minipage}
\, \hfill \,
\end{figure}
\hfill

\newpage

\noindent
\hfill
\begin{figure}[hbt]
\, \hfill \,
\begin{minipage}{0.7\linewidth}
\epsfig{file=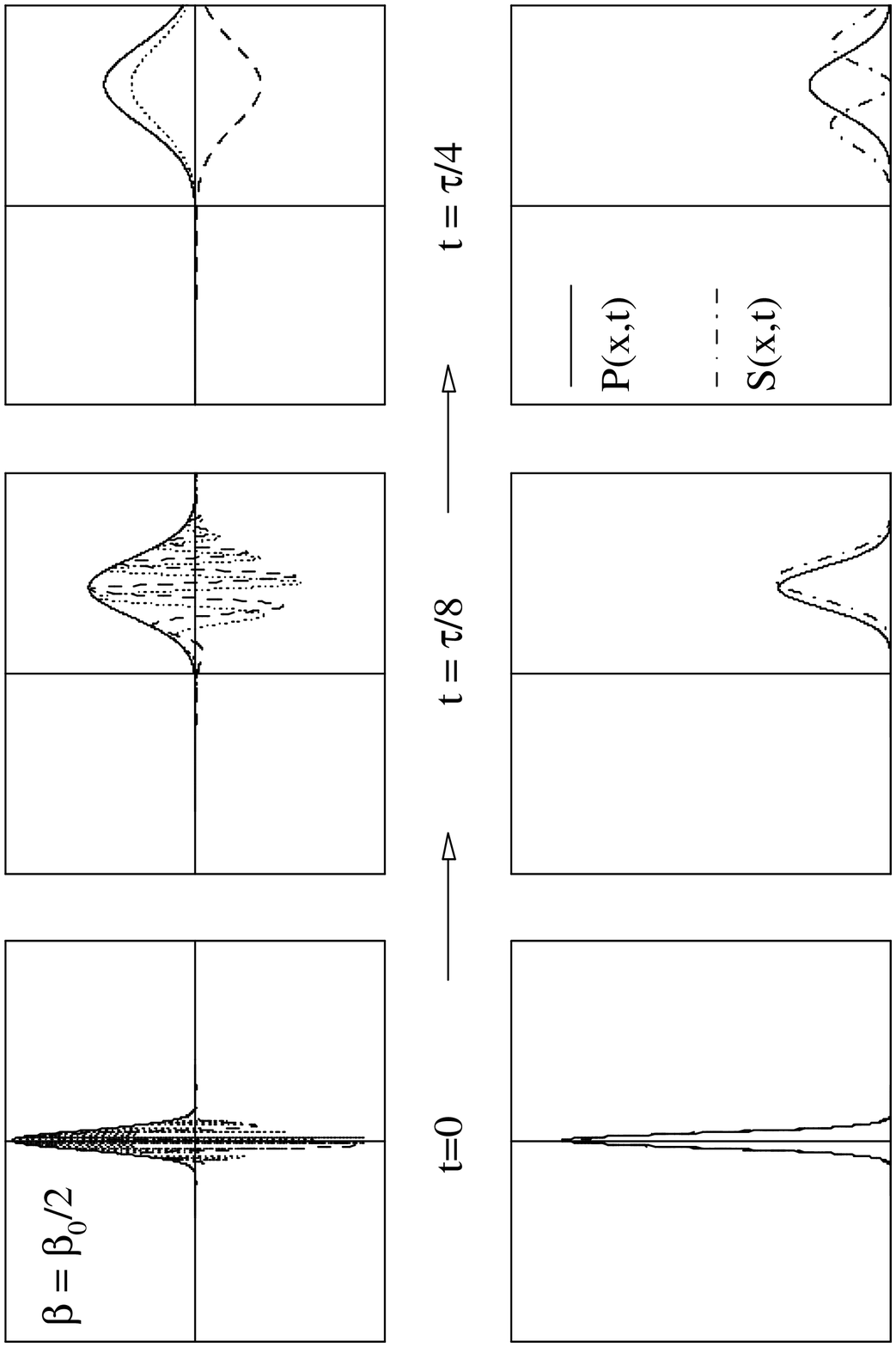,width=\linewidth}
\caption{}
\end{minipage}
\, \hfill \,
\end{figure}
\hfill

\newpage

\noindent
\hfill
\begin{figure}[hbt]
\, \hfill \,
\begin{minipage}{0.7\linewidth}
\epsfig{file=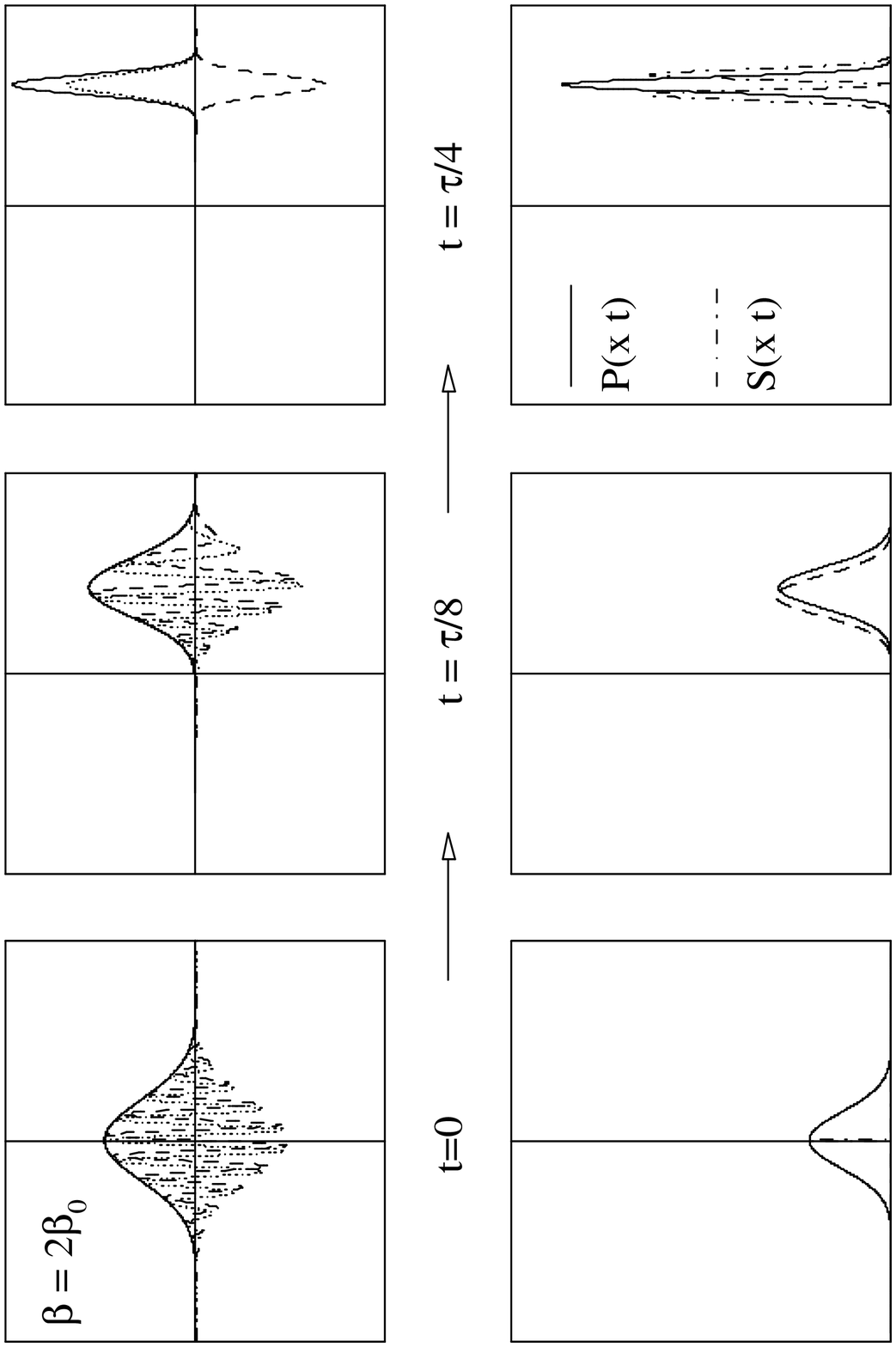,width=\linewidth}
\caption{}
\end{minipage}
\, \hfill \,
\end{figure}
\hfill

\newpage

\end{document}